\documentclass[twocolumn,aps,prl,showpacs,groupedaddress]{revtex4}
\usepackage{epsfig,amssymb,amsmath}

\def\comment#1{}\def\labell#1{\label{#1}}
\begin{document}
{\scriptsize Eprint: quant-ph/0510104}
\fbox{{\scriptsize Accepted version.}}
\title{Information-disturbance tradeoff in quantum measurements}
\author{Lorenzo Maccone}\affiliation{QUIT - Quantum Information Theory
  Group, Dipartimento di Fisica ``A.  Volta'' Universit\`a di Pavia,
  via A.  Bassi 6, I-27100 Pavia, Italy.}

\begin{abstract}
  We present a simple information-disturbance tradeoff relation valid
  for any general measurement apparatus: The disturbance between input
  and output states is lower bounded by the information the apparatus
  provides in distinguishing these two states.
\end{abstract}
\maketitle 

Extraction of information from a quantum system cannot always be
without feedback. This was clear since the early days of quantum
mechanics: It was the spirit of the original form of the Heisenberg
uncertainty ``principle'', as derived from the gedanken-experiment of
the Heisenberg microscope~\cite{heis}.  Since then, much more refined
descriptions of allowed quantum measurements have been put
forth~\cite{kraus}, so that we now know that the Heisenberg principle
can be easily circumvented~\cite{kraus,ozawaposition}, and that its
correct interpretation must be carefully adjusted (see
Ref.~\cite{mauro} for a recent review on the subject). The upshot is
that there is no ``unavoidable dynamical disturbance'' attached to all
measurements.  The debate on the uncertainty is no longer confined to
the realm of theory~\cite{kraus,ozawaposition,mauro,scully},
but experiments have also been carried out~\cite{rempe} confirming
that a feedback on the state of the system (due to information
extraction) is present even when all the possible dynamical
disturbances have been carefully eliminated. In a sense, this is to be
expected since the state of a system does not have a physical reality
{\it per se}, but it is a conceptual construct expressing the
information the experimenter has on the system~\cite{peres}. What is
most astonishing, is that such ``informational feedback'' {can} have
dynamical consequences: The subsequent evolution may drastically
change depending on the information extracted.

Is this feedback always present? If the initial state of the system is
known, then a measurement which extracts any kind of information
without changing the system state~\cite{kraus} is always
possible~\footnote{One may object that if the state is known, no
  information can be extracted: All possible information on the system
  is known through the knowledge of the state. This is true, however,
  only for pure states.}. Thus, it would seem that no
information-disturbance tradeoff relation can exist. In this paper,
however, we show that any informative measurement will affect at least
one state of the system. An information-disturbance tradeoff
concerning such a state can then be conceived: The amount of
disturbance on that state is lower bounded by the amount of
information that the measurement would return in distinguishing such
input from its corresponding output, see Eq.~(\ref{infodist}).

Various different information-disturbance tradeoffs have been proposed
previously~\cite{mauro,ozawanoise,peresfuchs,barnum,bana,bb84,decoy},
which explore different measures of information and of disturbance. In
this paper we use the most intuitive notions for these quantities:
Information is measured in bits through mutual information and
disturbance is measured using fidelity, which is the natural distance
measure for quantum states~\cite{jozsa,fuchsieee}.

In the following, we start by introducing the notation. We show that
at least one state must be modified by the measurement and then we
give a bound on such modification.  For the sake of clarity, we give
proofs of a very simple case, and postpone the general derivation to
the appendix.

Before attempting a derivation of an information-disturbance tradeoff,
we have to appropriately define these two quantities.  
\par\noindent
{\em Information:} Intuitively, one would expect that the information
extracted from a measurement should be defined as a function of the
outcome statistics only, such as the entropy of the probability of the
outcomes. This is easily shown to be inadequate: Think of a
measurement device that returns random outcomes (according to a well
defined probability) without yielding any information on the system. A
``good'' measurement should have outcomes in some way correlated to
the initial state of the system, so to provide information on the
system. Thus, a suitable expression for the information-part of our
tradeoff is through the mutual information $I$ the measurement
provides on which of two equally-probable input states the system is
in~\cite{mauro}. It supplies the fraction of a bit the measurement
tells us on which one is the input state, and varies continuously between
$I=0$ (no knowledge) and $I=1$ (complete knowledge). Alternatively, we
can employ the binary entropy $H_2(p_e)$ of the probability $p_e$ of
making an error when determining which state: It is a measure of the
uncertainty on the determination of which state. The two quantities
are simply related as $I=1-H_2(p_e)$. Information is measured in bits.
To obtain an adimensional quantity (in order to relate information and
disturbance), we will consider the ratio between information $I$ (or
uncertainty $H_2$) and the maximum information (or maximum
uncertainty) that can be obtained, i.e. one bit in this
case.\par\noindent {\em Disturbance:} A system is disturbed by a
physical process when its initial and final states do not coincide.
The fidelity $F(\varrho,\varrho')\equiv$Tr$[\sqrt
{\sqrt{\varrho}\varrho'\sqrt{\varrho}}]^2$~\cite{jozsa}, a simple
function of the Bures distance, is the most appropriate measure of the
``distance'' between the two states $\varrho$ and $\varrho'$. As such,
$1-F$ can be taken as a measurement of the disturbance~\cite{barnum}:
$1-F(\varrho,\varrho')=0$ if there is no disturbance (the output state
$\varrho'$ coincides with the input $\varrho$) and
$0<1-F(\varrho,\varrho')\leqslant 1$ if the input has been modified.
With this choice, a unitary evolution counts as a disturbing process,
even though it can be easily undone.  This might seem
unfortunate~\cite{mauro}, but a unitary evolution cannot provide any
information on the state, so its effect does not contrast the
information-disturbance tradeoff (according to which a disturbance
without information gain is possible). {\comment{Forse bisognerebbe
    espandere un po'... }}

Before deriving the tradeoff, we quickly review the necessary concepts
regarding quantum measurements. The postulates of quantum
mechanics~\cite{peres,kraus,ozawanoise} assert that the outcomes
statistics of any measurement is described by a POVM (Positive
Operator-Valued Measure), a set of positive operators $\{\Pi_k\}$
acting on the system Hilbert space $\cal H$ such that
$\sum_k\Pi_k=\openone$ ($\openone$ being the identity on $\cal H$):
The probability of the $k$th measurement outcome is
$p_k=$Tr$[\varrho\:\Pi_k]$, where $\varrho$ is the state of the system
prior to the measurement (Born rule). If the $k$th measurement outcome
occurred, the state evolves according to the following state-reduction
rule~\cite{kraus,krausprd,ozawanoise}
\begin{eqnarray}
\varrho'=\sum_{j\in I_k}K_j\:\varrho\:K_j^\dag/p_k
\;\labell{statered},
\end{eqnarray}
where the operators $K_j$ and the set of indices $I_k$ are such that
$\sum_{j\in I_k}K_j^\dag K_j=\Pi_k$. This implies that both the sets
$K_j$ and $U_jK_j$ (with arbitrary unitary operators $U_j$) give rise
to the same POVM $\{\Pi_k\}$ and thus to the same outcome statistics:
The post-measurement state is in general {\it not} determined by the
POVM elements. This is the reason why it is impossible to obtain an
information-disturbance tradeoff relation which is independent on the
system state. In fact, if we know the input state $\varrho$, we can
always tune the operators $U_j$ to reobtain the same state at the
output (if $\varrho$ is a mixed state, some additional classical
randomness might also be necessary).  For example, we can measure the
value of a qubit in the computational basis (using the POVM
$\{\Pi_0=|0\rangle\langle 0|,\;\Pi_1=|1\rangle\langle 1|\}$) and
always get as output state
$|+\rangle\equiv(|0\rangle+|1\rangle)/\sqrt{2}$, by choosing
$K_0=|+\rangle\langle 0|$ and $K_1=|+\rangle\langle 1|$. A striking
example of the same sort is a measurement of position which leaves a
particle in an eigenstate of the momentum~\cite{ozawaposition}.  The
physical interpretation of the operators $U_j$ is clarified by
considering a simple Stern-Gerlach measurement.  No sane
experimentalist who possesses a Stern-Gerlach apparatus oriented in
the $x$ direction rotates all his laboratory if he needs to measure a
$\frac 12$-spin along the $y$ axis.  He applies a unitary transformation to
rotate the spin with a magnetic field~\cite{krausprd}. In this case
the post-measurement state (if the spin is not absorbed) is an
eigenstate of $\sigma_x$, even though $\sigma_y$ was measured.

\begin{figure}[hbt]
\begin{center}
\epsfxsize=.75
\hsize\leavevmode\epsffile{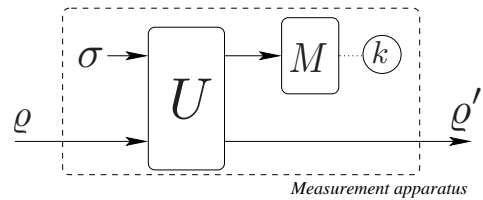}
\end{center}
\vspace{-.5cm}
\caption{Indirect measurement model. The system, initially in a state
  $\varrho$ impinges in the measuring apparatus (dashed line) which is
  initially prepared in the state $\sigma$. A unitary $U$ correlates
  the system and the apparatus. A projective measurement $M$ is then
  performed on the apparatus and yields the classical result $k$,
  which conditions the output state of the system $\varrho'$.}
\labell{f:imm}\end{figure}

Any evolution of the type~(\ref{statered}) can be derived from a
unitary evolution through the so-called indirect measurement
model~\cite{ozi,ozawanoise} (see Fig.~\ref{f:imm}).  The measured
system interacts unitarily with an external ancillary system
describing the measurement apparatus.  The ancillary system then
undergoes a L\"uders-type projective measurement $M$, i.e. such that
its POVM elements are orthogonal projectors $\{\Pi_k=|k\rangle\langle
k|\}$.  The system output state is then the partial trace (over the
ancillary Hilbert space $\cal A$) conditioned on obtaining the result
$k$ on the ancilla, i.e.~\cite{ozi,ozawanoise}
\begin{eqnarray}
\varrho'_{(k)}=\frac{\mbox{Tr}_{\cal A}
\left[(\openone_{\cal H}\otimes|k\rangle\langle
  k|)\:U\:(\varrho\otimes\sigma)\:U^\dag
\right]}{\mbox{Tr}
\left[(\openone_{\cal H}\otimes|k\rangle\langle
  k|)\:U\:(\varrho\otimes\sigma)\:U^\dag
\right]}
\;\labell{ozicollapse},
\end{eqnarray}
where $\sigma$ is the initial state of the ancilla and $U$ is the
unitary interaction that correlates the system to the apparatus,
acting on ${\cal H}\otimes{\cal A}$. Notice that there is no
assumption on the joint post-measurement state in
Eq.~(\ref{ozicollapse}), which combines the Born rule on the ancillary
space $\cal A$ with the rule to obtain the state of a subsystem from a
partial trace on the joint state.

For the sake of clarity, we will start analyzing the simple case in
which the input states of the system $\varrho$ and of the apparatus
$\sigma=|0\rangle\langle 0|$ are pure and no entanglement is generated
by the unitary $U$. The general situation will be analyzed
subsequently. The unitary will thus evolve two different input states
$|\psi_1\rangle$ and $|\psi_2\rangle$ according to the evolution
$|\psi'_1\rangle|a_1\rangle=U|\psi_1\rangle|0\rangle$ and
$|\psi'_2\rangle|a_2\rangle=U|\psi_2\rangle|0\rangle$. A unitary does
not change the scalar product, hence
$\langle\psi_1|\psi_2\rangle=\langle\psi_1'|\psi_2'\rangle\langle
a_1|a_2\rangle$. We assume that the measurement is informative, i.e.
the apparatus is able to correlate to the system somehow. This implies
that there must exist some $|\psi_1\rangle$ and $|\psi_2\rangle$ that
give rise to different states in the apparatus, i.e.
$|a_1\rangle\neq|a_2\rangle$. Thus, $|\:\langle a_1|a_2\rangle\:|<1$
so that
$|\:\langle\psi_1|\psi_2\rangle\:|<|\:\langle\psi_1'|\psi_2'\rangle\:|$,
i.e. the output states are less distinguishable than the input: their
fidelity has increased. In the general case (see the appendix), this
can be formalized in the following way. For any informative
measurement, there exist at least two system states $\varrho_1$ and
$\varrho_2$ such that
\begin{eqnarray}
F(\varrho_1,\varrho_2)<F(\varrho_1',\varrho'_2)
\;\labell{aumento},
\end{eqnarray}
where $\varrho'_1$, $\varrho'_2$ are the output states corresponding
to $\varrho_1$, $\varrho_2$ when the measurement results are the same.
This implies that for any measurement there exists at least one state
that is modified.

Call such a state $|\psi\rangle$.  The scalar product between
$|\psi\rangle$ and its evolved counterpart $|\psi'\rangle$ is
$|\:\langle\psi|\psi'\rangle\:|=|\:\langle\psi'|\psi''\rangle\langle
a|a'\rangle\:|\leqslant|\:\langle a|a'\rangle\:|$, where $|a\rangle$
and $|a'\rangle$ are the apparatus states corresponding to system
inputs $|\psi\rangle$ and $|\psi'\rangle$ respectively, and where
$|\psi''\rangle$ is the system output corresponding to input
$|\psi'\rangle$. [In general the evolution $U$ will generate
entanglement between system and apparatus so that the system output
state will be a mixed state (see appendix)]. The probability of error
$p_e$ in discriminating between two states $|a\rangle$ and
$|a'\rangle$ can be calculated from state discrimination
theory~\cite{helstrom} as $p_e=(1-\sqrt{1-|\:\langle
  a|a'\rangle\:|^2})/2$, whence $|\:\langle
a|a'\rangle\:|^2=4p_e(1-p_e)$. The uncertainty in this discrimination
is given by the Shannon entropy of the related probability
distribution $\{p_e,1-p_e\}$, i.e. the binary entropy $H_{2}(p_e)$. It
measures the bits of information one would gain by discovering which
of the two states the apparatus is in after the unitary interaction.
Since $4p_e(1-p_e)\leqslant H_2(p_e)$, we find that
$|\:\langle\psi|\psi'\rangle\:|^2\leqslant H_2(p_e)$: the fidelity
between the input and output states is upper bounded by the binary
entropy related to the discrimination of the two states by the
apparatus. This can be restated in the form of a tradeoff relation as
\begin{eqnarray}
1-|\:\langle\psi|\psi'\rangle\:|^2\geqslant 1-H_2(p_e)
\;\labell{tradeoff1}.
\end{eqnarray}
In the general situation (see the appendix), this
information-disturbance tradeoff takes the equivalent form
\begin{eqnarray}
1-F(\varrho,\varrho')\geqslant 1-H_2(p_e)
\;\labell{infodist}:
\end{eqnarray}
The disturbance $1-F$ between input $\varrho$ and output $\varrho'$ is
lower bounded by the mutual information $1-H_2(p_e)$ on which of the
two states $\varrho$ and $\varrho'$ is present at the input.  This is
the main result of the paper. By rearranging the terms
of~(\ref{infodist}) as $1-F(\varrho,\varrho')+H_2(p_e)\geqslant 1$, we
can also give it a different interpretation: The disturbance $1-F$
between input and output plus the uncertainty $H_2(p_e)$ in the
discrimination by the apparatus of these two states cannot be made
arbitrarily small. Equivalently, we can say that the mutual
information on which state plus the fidelity of these two states are
upper bounded by one.  

Since the inequality $4p_e(1-p_e)\leqslant H_2(p_e)$ is tight only for
$p_e=0$, $1/2$, and $1$, the bound~(\ref{infodist}) is not tight in
general. It is achieved only if the apparatus cannot discriminate
between $\varrho$ and $\varrho'$ at all, or if it can discriminate
between them exactly.

Even though the state reduction rule is not a quantum prerogative, the
tradeoff we derived is a purely quantum effect. In classical
mechanics, an informative non-disturbing measurement which perfectly
correlates the outcomes with the state of a system will collapse a
mixed state into a pure state: The effect of such a measurement is to
reduce the ``volume'' that the state of the system occupies in phase
space (a sort of ``classical state-reduction''). In classical
mechanics there is no lower bound to such volume and two pure states,
which occupy zero volume, can always be distinguished without
disturbance.  In contrast, in quantum mechanics the ``volume'' a state
must occupy in phase space is lower bounded by $\hbar/2$. On one hand
two non-identical pure states may overlap and their conclusive
discrimination may not be possible. On the other hand, if the
post-measurement state is perfectly correlated with the outcome
(L\"uders or von Neumann type apparatuses) and the measure is sharp
enough to sufficiently constrain the volume in one direction of the
phase space, the post-measurement state must ``expand'' in other
directions to preserve the minimum volume. For other types of
apparatuses the situation is not as clear-cut, but as we have shown,
at least one pure state of the system must be modified by any
informative measurement.  So, while in classical mechanics the system
will evolve compatibly with its pre-measurement trajectory in phase
space (only the ``thickness'' of the trajectory may be reduced), in
quantum mechanics the phase-space expansion might have observable
consequences and the system might not evolve compatibly with its
pre-measurement trajectory.

In conclusion, we have derived an information-disturbance tradeoff
which is valid for any measurement device: Any measurement modifies at
least one state of the system, and the fidelity between input and
output states is upper bounded by the information the apparatus is
able to extract when discriminating between input and output.  The
concept of conservation of quantum information~\cite{horod} was
inspirational: one can interpret the measurement as a correlation
between the initial state of the system and the measurement apparatus.

\section*{Appendix}
{\em Proof of Eq.~(\ref{aumento}):} Define the CP-map ${\cal L}_k$ as
the transformation described by the measurement with result $k$, see
Eq.~(\ref{statered}):   ${\cal L}_k(\varrho)\equiv\sum_{j\in
I_k}K_j\varrho K_j^\dag$. From the monotonicity of the fidelity under
maps~\cite{monot}, we know that $F(\varrho_1,\varrho_2)\leqslant
F({\cal L}_k(\varrho_1),{\cal L}_k(\varrho_2))$.  The equality holds
for any couple of input states $\varrho_1$, $\varrho_2$ only if the
map ${\cal L}_k$ is unitary~\cite{unit}, and such a map cannot convey
information on the system.  \comment{Dimostrazione di Vittorio (e'
  piu' elegante): If the equality holds for any couple of input states
  $\varrho_1$, $\varrho_2$, then it holds also for all couples of pure
  states $|\psi_1\rangle$ and $|\psi_2\rangle$. This, through Wigner's
  theorem~\cite{peres}, implies that the transformation is unitary,
  and such a map cannot convey information on the system.} In fact, a
unitary ${\cal L}_k$ on the system is obtained from a factorized
operator $U=U_{\cal S}\otimes U_{\cal A}$ in the indirect measurement
model of Eq.~(\ref{ozicollapse}). Any action on the system by such map
will be independent on the action on the probe, so that no information
on the system can reach the probe: The only maps which leave unchanged
the fidelity of any couple of input states are the unitaries, which
give no information. This can be stated equivalently in the following
manner. For any informative measurement, two states $\varrho_1$,
$\varrho_2$ exist such that Eq.~(\ref{aumento}) is true.

Incidentally, note that the converse also partially holds: If a
measurement decreases the fidelity, then all unitaries $U$
corresponding to its indirect measurement models will transfer some
information to the probe state (this does not automatically imply that
the measurement is informative, since the modification of the probe
state may be ignored the last stage of the apparatus, the von Neumann
measure $M$ of Fig.~\ref{f:imm}). In fact, the no-signaling property
of factorized unitary maps~\cite{nosig} implies that any
non-factorized unitary $U$ of the indirect measurement model can send
a signal from the system to the probe, i.e.  $U\neq U_{\cal H}\otimes
U_{\cal A}$ implies that there exist two states
$\varrho_1,\;\varrho_2$ such that $F(\sigma'_1,\sigma'_2)<1$, where
$\sigma'_i=$Tr$[U(\varrho_i\otimes\sigma)U^\dag]$ is the final state
of the probe, $\sigma$ is its initial state, and $U_{\cal H}$ and
$U_{\cal A}$ are arbitrary unitaries acting only on the system and on
the ancillary Hilbert spaces respectively.

It is possible to evaluate which states are modified by the
measurement process for each outcome $k$, by considering the map
${\cal L}_k$ as a linear operator on the operator space of the states
of the system.  One then immediately sees that only the eigenstates of
${\cal L}_k$ are not altered, while superpositions of eigenstates with
different eigenvalues are.

{\em Proof of Eq.~(\ref{infodist}):} In general, the input states to
the apparatus may be mixed. The probability of making a mistake when
discriminating two mixed states $\varrho_1$ and $\varrho_2$ is given
by
$p_e=1/2-\mbox{Tr}[\:|\varrho_1-\varrho_2|\:]/4$~\cite{max,fuchsieee}.
By using the property
Tr$[\:|\varrho_1-\varrho_2|\:]/2\leqslant\sqrt{1-F(\varrho_1,\varrho_2)}$~\cite{fuchsieee},
we can write $p_e\geqslant(1-\sqrt{1-F(\varrho_1,\varrho_2)})/2$,
where the equality is attained for pure states~\cite{helstrom}. The
binary entropy $H_2(x)\equiv -x\log_2x-(1-x)\log_2(1-x)$ for
$x\in[0,1]$ satisfies the inequalities $x\leqslant
H_2(1/2-1/2\sqrt{1-x})$ and $x\leqslant H_2(1/2+1/2\sqrt{1-x})$.
Moreover, for $x\leqslant 1/2$, it is monotonically increasing so that
we can write
\begin{eqnarray}
x\leqslant H_2\Big(\frac 12-\frac 12\sqrt{1-x}\Big)\leqslant H_2(y)
\;\labell{disug1},
\end{eqnarray}
for any $y$ such that $\frac 12-\frac 12\sqrt{1-x}\leqslant y\leqslant
\frac 12$. Choosing $x=F(\varrho_1,\varrho_2)$ and $y=p_e$, we obtain
$F(\varrho_1,\varrho_2)\leqslant H_2(p_e)$, i.e.  Eq.~(\ref{infodist})
from~(\ref{disug1}), which is valid when $p_e\leqslant 1/2$. If
$p_e\geqslant 1/2$ instead, we proceed analogously starting from
\begin{eqnarray}
x\leqslant H_2\Big(\frac 12+\frac 12\sqrt{1-x}\Big)\leqslant H_2(y')
\;\labell{disug2},
\end{eqnarray}
valid for $1/2\leqslant y'\leqslant\frac 12+\frac 12\sqrt{1-x}$.
Choosing $x=F(\varrho_1,\varrho_2)$ and $y'=1-p_e$, we obtain
Eq.~(\ref{infodist}) for $p_e\geqslant 1/2$, by recalling that
$H_2(1-p_e)=H_2(p_e)$.

\begin{acknowledgments}
  I thank Vittorio Giovannetti for very useful discussions and
  criticisms.  Financial support comes from the Ministero Italiano
  dell'Universit\`a e della Ricerca (MIUR) through FIRB (bando 2001)
  and PRIN 2005.
\end{acknowledgments}

\end{document}